# Controlling selenization equilibrium enables high-quality Cu$_2$ZnSn(S, Se)$_4$ absorbers for efficient solar cells

Xiao Xu†, Jiazheng Zhou†, Kang Yin, Jinlin Wang, Licheng Lou, Menghan Jiao, Bowen Zhang, Dongmei Li, Jiangjian Shi*, Huijue Wu, Yanhong Luo*, and Qingbo Meng*

* Corresponding authors.
† These authors contributed equally.

X. Xu, J. Zhou, K. Yin, J. Wang, L. Lou, M. Jiao, B. Zhang, D. Li, J. Shi, H. Wu, Y. Luo, Q. Meng
Beijing National Laboratory for Condensed Matter Physics, Institute of Physics, Chinese Academy of Sciences (CAS), Beijing 100190, P. R. China
E-mail: qbmeng@iphy.ac.cn; yhluo@iphy.ac.cn; shijj@iphy.ac.cn

Q. Meng
Center of Materials Science and Optoelectronics Engineering, University of Chinese Academy of Sciences, Beijing 100049, P. R. China

X. Xu, J. Zhou, K. Yin, J. Wang, L. Lou, M. Jiao, B. Zhang, D. Li, Y. Luo
School of Physical Sciences, University of Chinese Academy of Sciences, Beijing 100049, P. R. China

D. Li, Y. Luo, Q. Meng
Songshan Lake Materials Laboratory, Dongguan, Guangdong 523808, P. R. China

**Abstract**

$Cu_2ZnSn(S, Se)_4$ (CZTSSe) is one of most competitive photovoltaic materials for its earth-abundant reserves, environmental friendliness, and high stability.[1-3] The quality of CZTSSe absorber determines the power-conversion efficiency (PCE) of CZTSSe solar cells. The absorber's quality lies on post-selenization process, which is the reaction of Cu-Zn-Sn precursor and selenium vapor. And the post-selenization is dependent on various factors (e.g. temperature, precursor composition, reaction atmosphere, etc).[4-6] However, synergistic regulation of these factors cannot be realized under a widely-used single-temperature zone selenization condition.[7,8] Here, in our dual-temperature zone selenization scheme, a solid-liquid and solid-gas (solid precursor and liquid/gas phase Se) synergistic reaction strategy has been developed to precisely regulate the selenization. Pre-deposited excess liquid Se provides high Se chemical potential to drive a direct and fast formation of the CZTSSe phase, significantly reducing the amount of binary and ternary compounds within phase evolution. And organics removal can be accomplished via a synergistic optimization of Se condensation and subsequent volatilization. We achieve a high-performance CZTSSe solar cell with a remarkable PCE of 13.6%, and the highest large-area PCE of 12.0% (over 1cm$^2$). Our strategy will provide a new idea for further improving efficiency of CZTSSe solar cells via phase evolution regulation, and also for other complicated multi-compound synthesis.

CZTSSe has the advantages of high light absorption coefficient, earth-abundant reserves, adjustable band gap, environmental friendliness and high stability, etc.[1-3,9] It is believed to be one of best material for thin-film solar cells, and the theoretical limit efficiency of CZTSSe solar cells is as high as 32%.[10] In particular, the CZTSSe solar cell based on the solution route is easier to realize mass production and can further reduce costs, and is the most potential and most valuable new thin-film solar cell. After years of efforts, the solvents used in CZTSSe solar cells have changed from highly toxic hydrazine to environmentally friendly green solvents,[11-15] and the record efficiency has increased from 12.6% to 13.0%.[16,17] However, there is still a huge gap between the current efficiency and the theoretical efficiency. And this difference is mainly attributed to high loss of open-circuit voltage, which originates from the poor crystal quality and various types of defects.[4,18-21]

The challenges in preparing high-quality CZTSSe crystals lie in their diverse constituent elements, narrow phase diagram, and complex crystallization processes.[5,22,23] First, from the perspective of chemical reactions, the crystallization process of CZTSSe is the solid-gas and solid-liquid chemical reactions between the precursor film and selenium in a high-temperature environment. These reactions are affected by the initial Se content, the concentration and uniformity of selenium vapor, reaction temperature, etc.[8,24] Second, from a thermodynamic point of view, the reaction between Cu-Zn-Sn precursor and Se easily forms binary or ternary phases with lower Gibbs free energy than CZTSSe,[5] and these phases will lead to a large amount of defects in the final film.[6,25] Third, from a kinetic perspective, each cation has different diffusion rate and volatilization rate during the crystal growth, resulting in extra element loss in the absorber.[26-28] Thus, there are many reaction conditions in the crystallization process (such as precursor composition, reaction temperature, reaction time, reaction atmosphere, Se concentration, and uniformity of reactants, etc.) affecting the crystal quality of the absorber.[22,29,30] Precisely controlling and optimizing these reaction kinetic parameters is the key to high-quality absorber.

The main aspects to improve the CZTSSe crystal growth cover designing the precursor composition and optimizing the selenization reaction process. Considering the precursor regulation, by changing the valence state of precursor cations,[4] improving local chemical environment,[6] and cation doping,[31,32] etc., the segregation of the secondary phase in the reaction evolution and Sn-related defects are suppressed to a certain extent. In addition, by introducing appropriate Se into the precursor (dissolving elemental selenium in the solution or thermally evaporating elemental selenium on the surface of the precursor film, etc.), the initial concentration of selenium in the nucleation stage of the selenization reaction is increased, thereby mitigating the selenium-deficient atmosphere and lowering Se-related defect density.[24,33] Compared to precursor regulation, the optimization of selenization reaction process is relatively less focused on. There are two limitations in the widely used single-zone graphite-box selenization. First, as a narrow and closed space, the single-zone graphite box fails to realize independent control of reaction temperature, vapor concentration and the initial amount of Se content. Second, due to the characteristics of volatile selenium and strong penetration in graphite, it is difficult to solve the problem of selenium deficiency in the crystal growth/ ripening stage, even with the initial selenium content increasing.[8] Therefore, it is urgent to develop new selenization technology to precisely control the kinetic process of selenization reaction, that is, it can take into account both introducing selenium source and building up the selenium balance between reactants and reaction atmosphere, and realize the synergistic control of various reaction parameters.

In this paper, we developed a solid-liquid and solid-gas synergistic reaction (SLSG) strategy by adopting dual-temperature zone scheme. Herein, sufficient liquid Se is pre-deposited on the surface of the precursor film to achieve liquid-phase assisted growth. The benefits of this strategy can be summarized as below. First, the liquid Se provides a high chemical potential to drive a faster direct phase transformation process of CZTSSe in the initial stage of selenization. Second, the high concentration of Se also suppresses the surface decomposition and element valence state variation of CZTSSe. Third, the crystallization of CZTSSe and the organics removal can be well balanced via synergistically controlling the selenium volatilization process. These advantages help us to realize defect-less and compact CZTSSe absorbers, which contribute to a high-performance solar cell with an efficiency of 13.6% and a large-area (over 1 cm$^2$) device with the highest efficiency of 12.0%. These results are among the highest reported to date.

**Key issues in realizing high-quality CZTSSe**

Figure 1(a) gives the phase diagrams of CZTSe and secondary phases. We can see that phase evolution process strongly depends on vapor Se concentration ($P$) and the reaction temperature ($T$).[5] The low Se concentration and temperature is usually favorable for the preferential formation of $Cu_2Se$, $ZnSe$, $Cu_2SnSe_3$ and $SnSe_2$, while the direct formation of CZTSSe requires more concentrated Se vapor and much higher temperature. However, uniform and sufficient Se atmosphere is usually difficult to realize, especially during the initial few minutes of selenization. First, we simulate the behaviors of Se volatilization and spatial diffusion in single-zone graphite-box. The results (Figure S1) shows that Se vapor within graphite-box is apparently non-uniform and its concentration is much lower compared to saturated vapor pressure. Moreover, within one closed space in graphite box, the reaction factors like reaction temperature, vapor concentration and the initial amounts of Se reactants, strongly influence each other. Thus, the graphite-box selenization route may bring the multi-step phase evolution and their induced defects are almost inevitable and out of control.

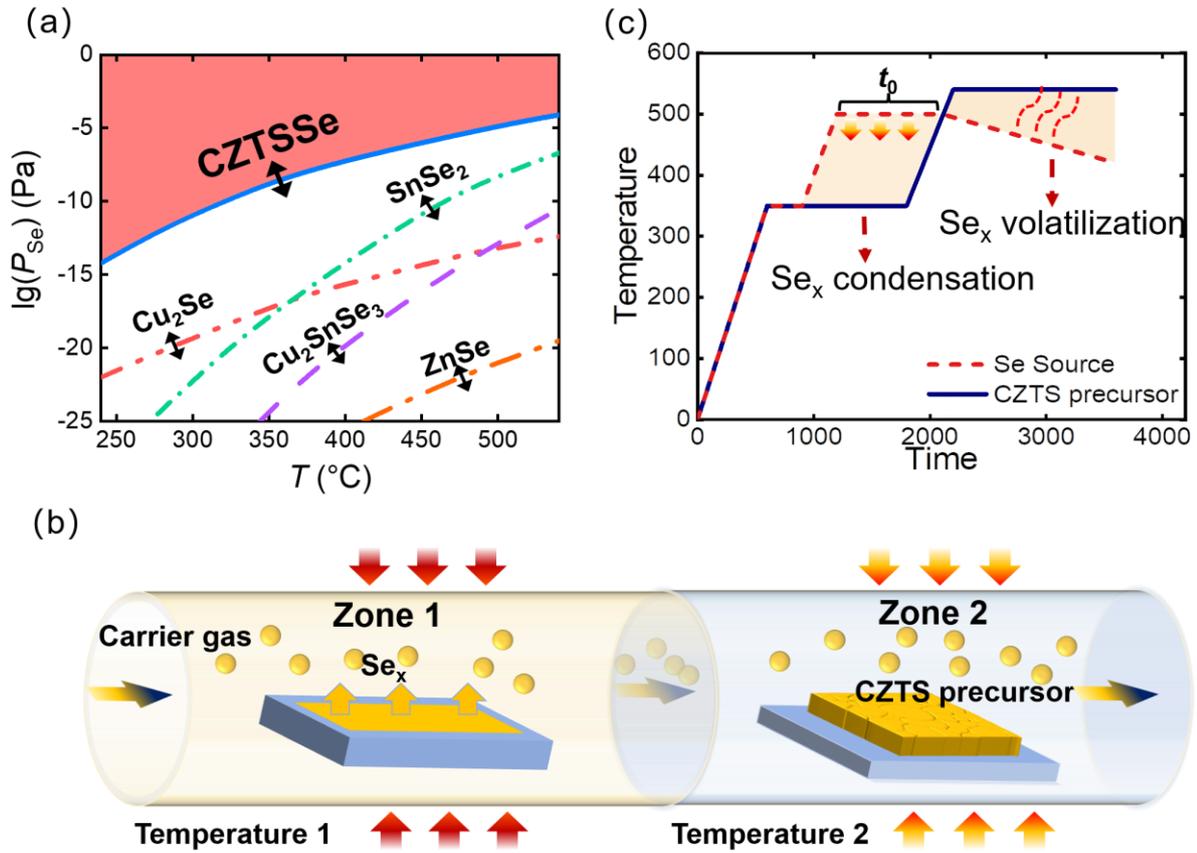

Figure 1. (a) Schematic Se vapor equilibrium versus temperature for the formation reactions of CZTSSe and secondary phases.[5,27] (b) The schematic of dual-temperature zone selenization system. The Se$_x$ vapor generated from Zone 1 is transferred into Zone 2 via carrier gas, and reacts with CZTS precursor. (c) Typical time-dependent temperature evolutions of CZTS precursor and Se source designed in this work. A condensation process and a volatilization process are built up through design of temperature difference and time difference. $t_0$ represents the time delay of initial heating time of Se source and CZTS precursor.

**Solid-liquid and solid-gas synergetic reaction strategy**

Aiming at the multi-parameter coupling problem, we take a decoupling strategy via space-time separation. Figure 1(b) shows our designed dual-temperature zone selenization approach. In this system, the CZTS precursor and Se source are spatially separated and their heating programs can be controlled independently. The molecular Se is transported from Se source to CZTS precursor via carrier gas. Owing to this separation of reactants (CZTS precursor and Se source) and their heating programs, more reaction pathways can be explored.

Inspired by liquid phase epitaxy in Si[34], SiC[35] and III-V semiconductor[36], we select the liquid-phase Se to provide orders of magnitude higher molecular Se concentration than Se vapor in the initial stage of selenization. The introduction of liquid Se is realized by building up temperature difference between selenium source and CZTS precursor. The amount of liquid Se is proportional to the temperature difference and its duration. First, the Se source is preheated to high temperature, and the gaseous Se is transported to CZTS precursor. Second, due to the much lower temperature in the Zone 2, the supersaturated Se vapor will condense onto the surface of CZTS precursor to form a liquid state. Third, when the temperature of precursor is increased to the selenization reaction started, the liquid Se will react with the solid CZTS precursor directly. More importantly, to extend the reaction time of liquid Se and CZTS precursor, it is necessary to provide a continuous and appropriate Se vapor pressure from Se source.

**Synergetic optimization in nucleation and growth stage**

To be specific, we keep the temperature difference constant and vary the duration of the preheating stage (as marked with $t_0$ in Figure 1(c)) to determinate the amount of the liquid Se in the precursor film, and compare the solid-gas selenization and selenization involving solid-liquid and solid-gas reaction. According to the optical microscope images (Figure S3), there is no liquid-phase Se found on precursor surface in the condition of $t_0 = 0$ s. When $t_0$ reaches 300 s, the precursor is entirely covered by liquid-phase Se. Here, we define the condition $t_0 = 0$ s as pure solid-gas selenization (SG) and the condition $t_0 = 300$ s as the solid-liquid and solid-gas synergetic selenization (SLSG). Figure 2(a1)-(a2) show top-view images (under SG and SLSG route, respectively) of semi-selenized films by interrupting selenization when the CZTSSe precursor reaches 540°C and maintains for 200 s. In SLSG route, there is a large amount of cooled liquid Se distributed on the surface and in the grain boundaries, suggesting the liquid-phase Se assisted growth mechanism has been successfully proposed. The influence of liquid-phase Se on the phase formation processes are studied by Raman and X-ray diffraction (XRD) characterization. The film in the intermediate stages for this study is sampled by interrupting the selenization when the temperature of CZTS precursor reaches 400 and 500 °C, respectively. As shown in Figure 2(b1)- (c1), when the selenization adopts a conventional solid-gas reaction route, although the CZTSSe phase has already appeared, the $Cu_xSe$, $Cu_2SnSe_3$ and SnSe secondary phases indicated by the Raman peaks of 265, 180 and 130 cm$^{-1}$ are clearly seen [37,38]. Specifically, the $Cu_xSe$ binary phase is firstly formed at 400 °C, then transformed into the $Cu_2SnSe_3$ phase at 500 °C. In the SLSG route, these secondary phases in the intermediate selenization process are completely eliminated, as shown in Figure 2(b2)-(c2). To be noticed, even if the precursor is half covered by liquid Se, the suppression of secondary phase is also apparent (Figure S4). In addition, the XRD characterizations (Figure S5) suggest relatively faster phase transformation from precursor to CZTSSe in SLSG. These results mean that the extremely high molecular Se concentration from liquid Se has realized a direct and fast formation of the CZTSSe phase.

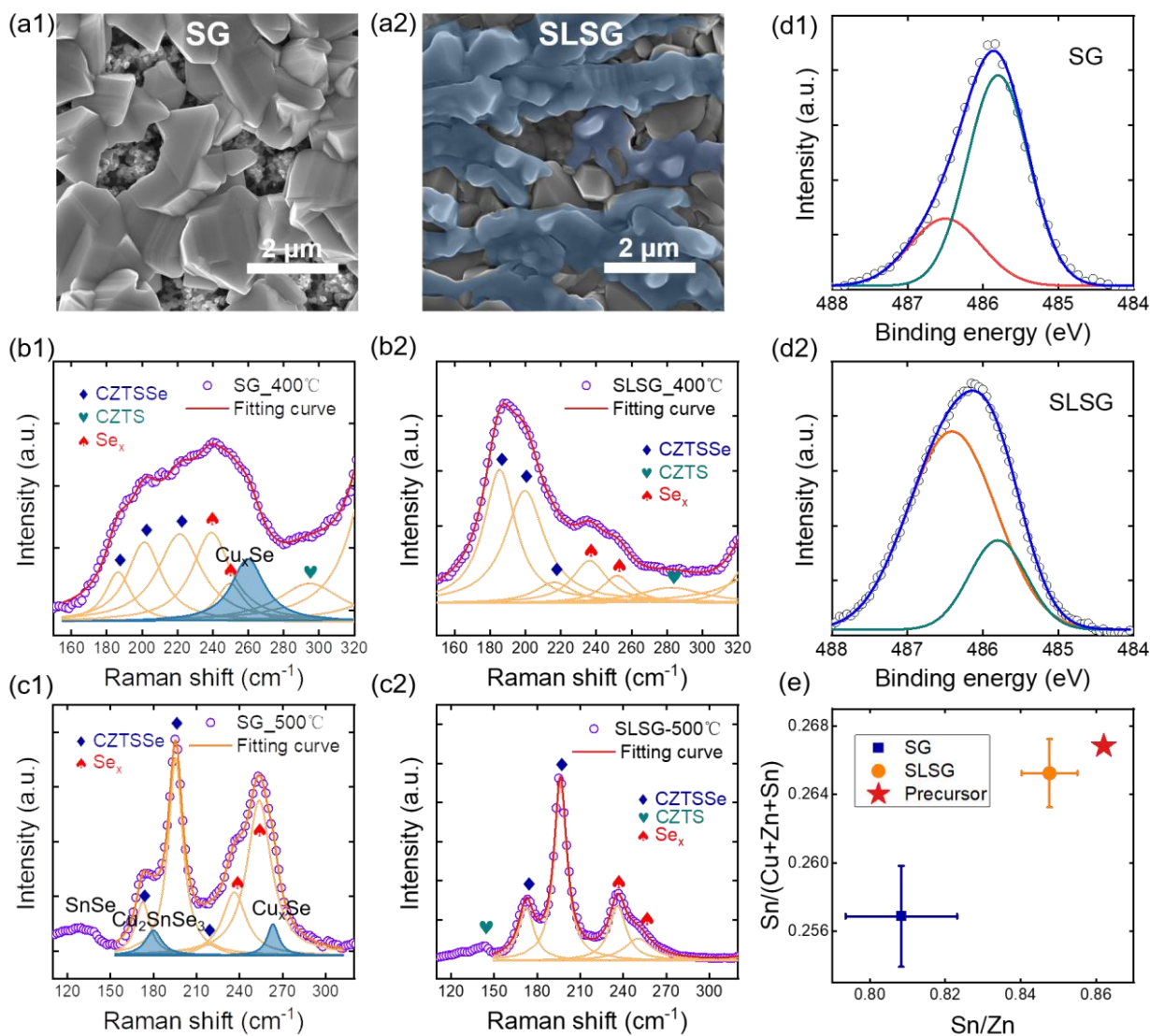

Figure 2. Top-view SEM images of selenized films based on (a1) solid-vapor reaction and (a2) solid-liquid reaction. The bule region is the cooled liquid Se. Raman spectra of (b1) SG-400 °C, (b2) SLSG-400 °C, (c1) SG-500 °C and (c2) SLSG-500 °C. The circles represent raw data and red lines give the peak fitting results. Fitting peaks of CZTSSe, CZTS and Se$_x$ are labelled by diamond, heart, spade, respectively; The color-filled peaks at 180 cm$^{-1}$ and 265 cm$^{-1}$ belong to Cu$_2$SnSe$_3$ and Cu$_x$Se. The X-ray photoelectron spectra (XPS) spectra of (d1) SG and (d2) SLSG. Circles show the raw data, and lines give the muti-peak fitting results. Orange lines represent Sn$^{4+}$ and green lines represent Sn$^{2+}$. (e) The energy dispersive X-Ray fluorescence (XRF-EDX) results of SG sample and SLSG

sample. And the red star represents the precursor.

We further compare chemical environments and composition of SG sample and SLSG sample via X-ray photoelectron spectra (XPS) and energy dispersive X-Ray fluorescence (XRF-EDX) spectra. In SG sample, the Sn ($3d^{5/2}$) peak is located at 485.9 eV with an asymmetric shape as in Figure 2(d1). The main component of this peak is fitted to be at 485.8 eV, corresponding to the $Sn^{2+}$ cation in the SnSe [6,39]. This implies that the CZTSSe surface has experienced decomposition reaction, for example, CZTSSe → $Cu_2Se$ + ZnSe + SnSe + Se[5,27]. The decomposition of the intermediate $SnSe_2$ could also form the $Sn^{2+}$ cation. The decomposition product SnSe is highly volatile at high temperature, which will cause Sn loss in the film.[27] Our XRF characterization clearly confirms this result (Figure 3e). The Sn loss or insufficient Se atmosphere will induce the Sn vacancy in the CZTSSe. Both these two types of atomic vacancies have been shown to be deep defects in this material[18,22]. Differently, in SLSG sample, the main component of the $Sn3d^{5/2}$ peak is located at 486.4 eV, corresponding to the $Sn^{4+}$ cation, indicating that the surface decomposition has been suppressed (Figure 3(b)).[40] The element composition measurement also demonstrates that the Sn loss has been obviously reduced. Obviously, this SLSG route is able to suppress the appearance of the Sn vacancy deep defect. We also find that the liquid Se can improve the nucleation and morphology evolution of the selenized film (Figure S6). All the above characterizations prove that solid-liquid reaction is favorable for high-quality CZTSSe crystal, as shown in Figure S7.

**Synergetic optimization in ripening stage**

In environment-friendly solution routes, another challenge is how to balance the crystal ripening and organics removal, because both two processes are sensitive to the concentration of Se vapor. Even though better CZTSSe crystals have been realized via SLSG route, corresponding solar cells present much lower PCE than that of SG (Figure 3(a)), and mainly drop in $J_{SC}$ and FF. This suggests poor charge transportation ability, which is mainly attributed to excess organics. Addressing this, we further regulated SLSG route via optimizing the declining rate of Se concentration in ripening stage. And this optimized SLSG process is labelled as SLSG-O.

Cross morphology and composition characterizations of SLSG samples and SLSG-O samples are performed. As in Figure 3(b), the SLSG sample presents a double-layer structure with a top large-grain layer and a bottom amorphous layer. FTIR results (Figure 3(c)) show that this amorphous layer in SLSG sample is a kind of C-N framework, for the occurrence of a broad IR peak at 1100~1300 cm$^{-1}$. Furthermore, no graphite signals can be found in Raman spectra (Figure S9). Thus, this C-N framework is insulative, which will remarkably influence charge transportation, unlike the graphite-like carbon framework we reported before.[41] In SLSG-O samples, however, the signal of C-N bonding disappears in FTIR spectra, indicating the complete removal of insulative organic residue. And benefitting from the breakdown of C-N framework, the crystals in the bottom layer can be well-fused into large grains (Figure 3(b)). The Kelvin probe force microscopy (KPFM) is performed to study the contact-potential difference (CPD) of SLSG and SLSG-O samples (Figure 3(d1-2)). The results show that, the uniformity of CPD has been significantly improved and the average CPD of CZTSSe film decreases from 460 to 100 mV in the SLSG-O sample. This means the p-type doping of CZTSSe has been enhanced. Furthermore, the large negative signal in transient photocurrent (TPC) spectra under 0.5 V (Figure 3(e)) disappears via controlling volatilization of Se, suggesting the barrier of charge transportation in SLSG-O sample has been successfully removed.[42] Finally, the average PCE of SLSG-O devices are significantly improved (Figure 3(f1-4)) from 10 % to 12.6 % and the highest PCE is 13.1%. The highest open-circuit voltage reaches 558 mV, owing to the better crystal-quality. And the improved average FF of 0.68 is attributed to better charge transportation enabled by the complete removal of insulate organics.

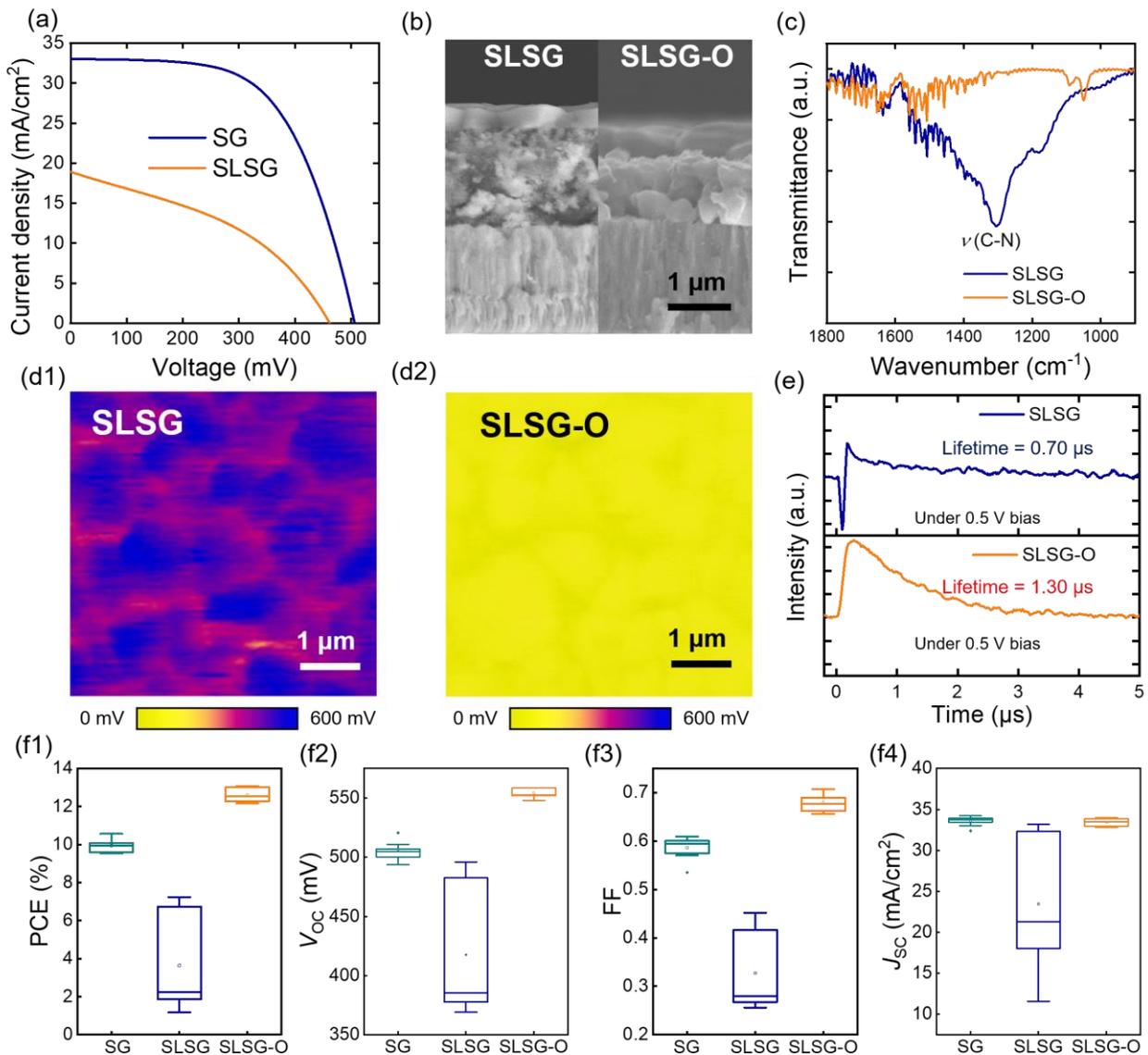

Figure 3 (a) *J-V* curves of SG and SLSG. (b) The cross-section SEM images of SLSG and SLSG-O selenized films. (c) The FTIR spectra of the SLSG sample and SLSG-O sample. The broad peak ranging from 1100 to 1300 cm$^{-1}$ in SLSG, belongs to C-N bonding, which suggests the existence of organic residues. The surface potential distributions (SPD) of (d1) SLSG and (d2) SLSG-O. The surface potential of SLSG-O sample is 400 mV lower than that in SLSG sample, suggesting better conductivity. (e) The TPC curves under 0.5V bias of SLSG and SLSG-O. The device performance of SG, SLSG and SLSG-O: (f1) PCE, (f2) $V_{OC}$, (f3) FF, (f4) $J_{SC}$.

**Defect properties and device performance**

We further use modulated electrical transient measurements to study the influence of selenization routes on the final cells. According to modulated transient photocurrent (m-TPC) measured at -1 V in Figure 4(a1-2, b1-2), the decay of the SLSG-O is much faster than that of SG, in addition, the time position of the TPC peak for the SG sample is almost independent to the applied voltage, while the rise of the TPC signal of the SLSG-O sample is obviously slowed down when increasing the voltage.[43] And a better collection efficiency can be seen in SLSG-O (Figure 4(c)).[44] These phenomena indicate that the SLSG-O sample has a much better carrier transport ability, mainly benefiting from the better crystallization quality, less secondary phases and carrier trapping states[44]. Obvious difference in carrier recombination properties between these two samples is also observed from the modulated transient photovoltage (m-TPV) results. For the SG sample, the TPV measured at 0 V exhibits a dual-exponential decay dynamics, with a fast decay in the early stage. This fast decay is caused by the carrier recombination in the CZTSSe surface, induced by decomposed surface phases or related defects. For the SLSG-O sample, its photovoltage decay exhibits a single exponential dynamic behavior with much longer lifetime, corresponding to remarkable suppression of carrier recombination in the cell. Furthermore, TPV lifetime of SLSG-O has much stronger dependence to the voltage and better performance in resulting ideal factor than that of SG (Figure S10). Obviously, SLSG-O route can well improve heterojunction properties of the cell. And other electrical characterizations (DLCP and *C-V*) also confirm this result (Figure 4(d) & S11). The interface defect density in the SG sample estimated from the difference between the DLCP and the *C-V* results is over 15 times larger than that of the SLSG-O sample[45]. Moreover, suppression to the carrier nonradiative recombination in the SLSG-O sample is also supported by temperature-dependent photoluminescence (PL) of the CZTSSe film, which exhibit much higher intensity (Figure 4(e)) and an obvious PL bule shift of about 40 meV (Figure S12).[46]

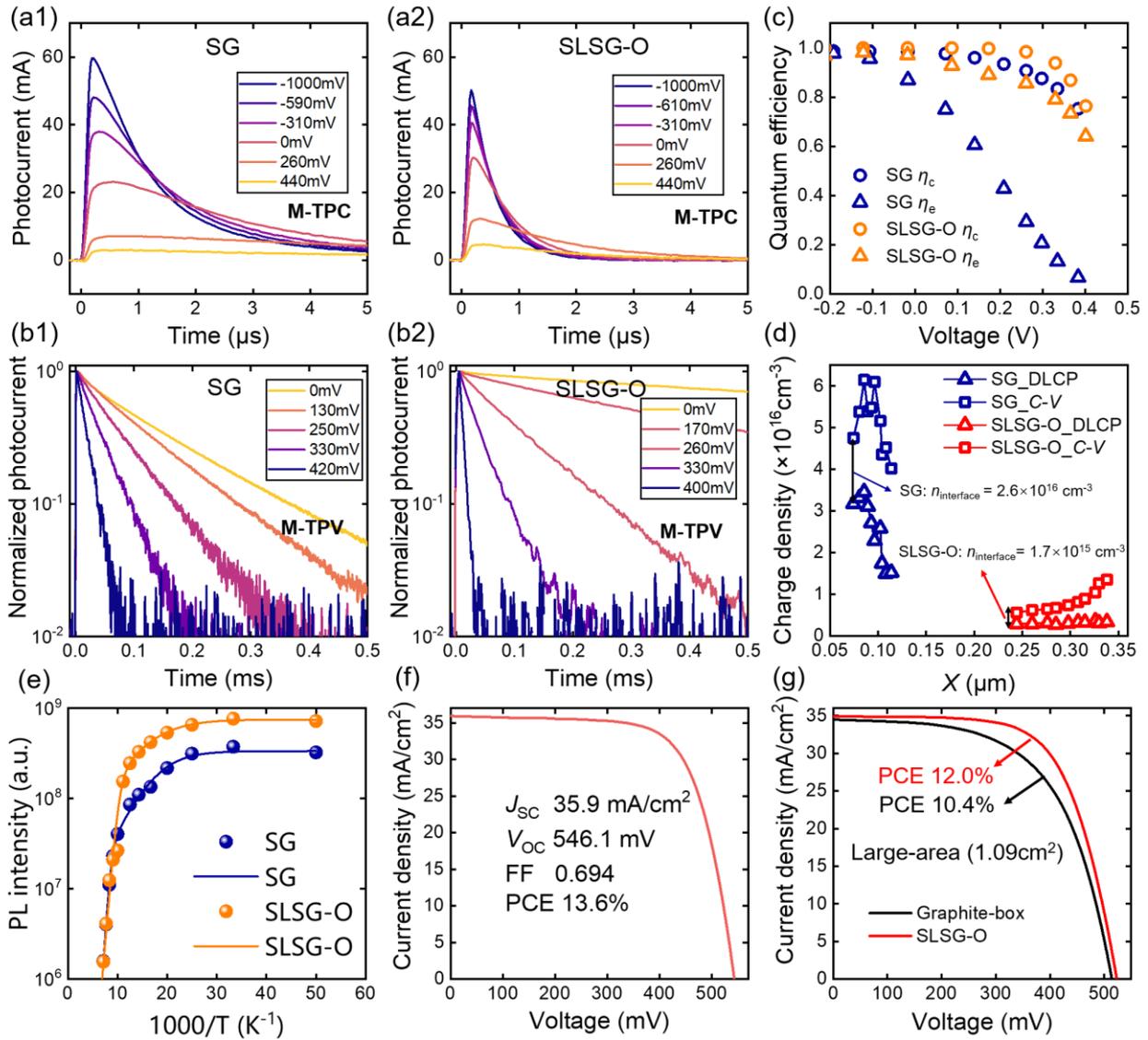

Figure 4. M-TPC spectra of (a1) SG and (a2) SLSG-O. The M-TPV spectra of (b1) SG and (b2) SLSG-O. (c) The collection efficiencies $\eta_c$ and extraction efficiencies $\eta_e$ of SSG and SLSG-O. (d) The *C-V* and DLCP characterizations of SG and SLSG-O. The interface defects are successfully suppressed in SLSG-O. (e) Integral PL intensity versus $1000/T$ curves. Dots are the raw data and lines are fitting curves. SLSG-O has higher PL intensity and defect activation energy. (f) *J-V* curve of device with 13.6% PCE. (g) The *J-V* curves of large-area (1.09 cm$^2$) solar cells basing on graphite box selenization (black line) and SLSG-O (red line)

Finally, a high-performance cell with a total-area efficiency of 13.6% (certified 13.4%, as in Figure S13) was achieved, which are amongst the highest results reported for CZTSSe solar cells. The current-voltage (*J-V*) characteristics of the cell is shown in Figure 4(f) and the external quantum efficiency spectrum (EQE) is given in Figure S14. And The derived $E_g$ is 1.10 eV and the integral short-circuit current density is 37.0 mA/cm$^2$, quite close to the *J-V* result. The detailed device performance parameters of our cell and state-of-the-art CZTSSe solar cells are summarized in Table 1. Our cell exhibits the $V_{OC}$ of 546.1 mV and the corresponding $V_{OC}$ deficit ($E_g/q$-$V_{OC}$) is 0.554 V. This $V_{OC}$ deficit value is obviously lower than those of the 12.6% and 13.0% record cells[16,17]. Furthermore, compared to graphite-box selenization, our strategy demonstrates more impressive performance in large-area (1.09 cm$^2$) device, resulting a remarkable PCE of 12.0% (Figure 4(g)).

Table 1. Device performance parameters of efficient kesterite solar cells

| Device | PCE(%) | $J_{SC}$ (mA/cm$^2$) | $V_{OC}$ (mV) | FF | $E_g$ (eV) | $E_g/q$-$V_{OC}$ (V) |
|---|---|---|---|---|---|---|
| IBM cell[16] | 12.6 | 35.2 | 513 | 0.698 | 1.13 | 0.617 |
| DIGST cell[30] | 12.6 | 35.4 | 541 | 0.659 | 1.13 | 0.589 |
| NJUT cell[17] | 13.0 | 33.6 | 529 | 0.729 | 1.11 | 0.581 |
| This work | 13.6 | 35.9 | 546.1 | 0.694 | 1.10 | 0.554 |

**Conclusions**

In this work, we adopted a dual-temperature zone selenization scheme to realize a solid-liquid and solid-gas synergistic selenization reaction strategy. The large amount of liquid Se has induced a solid-liquid reaction pathway and the high Se chemical potential realizes a direct and fast formation of the CZTSSe phase. In the subsequent stage, a synergetic regulation of Se condensation and volatilization enables both better crystal growth and organics removal. Finally, the CZTSSe films with low bulk and surface defects have been achieved, which contributes to a remarkable device PCE of 13.6% and large-area (1.09 cm$^2$) PCE of 12.0%. Overall, this work explores a promising direction to precisely control the selenization, and its reaction mechanism will have great referential significance to other complicated multi-compound synthesis.